# Signatures of Kitaev quantum spin liquid states in α−$RuCl_3$ revealed by nitrogen-vacancy spin-noise spectroscopy

Jitender Kumar,[1*] Sehrish Iqbal,[1] Himani Saraswat,[1] Benjamin Hammons,[1] Tianlin Li,[2] Ariel Smooha,[5] Rupak Timalsina,[1] Arnab K. Pariari,[3,4] Sy-Hwang Liou,[2] Xia Hong,[2] Markus Hücker,[3] Amit Finkler,[5*] and Abdelghani Laraoui[1,2*]

[1]Department of Mechanical & Materials Engineering, University of Nebraska-Lincoln, Lincoln, NE 68588, United States

[2]Department of Physics and Astronomy and Nebraska Center for Materials and Nanoscience, University of Nebraska-Lincoln, Lincoln, NE 68588, United States

[3]Department of Condensed Matter Physics, Weizmann Institute of Science, Rehovot 7610001, Israel

[4]Quantum Technology Institute (QuTI), SRM University-AP, Amravati 522240, Andhra Pradesh, India

[5]Department of Chemical and Biological Physics, Weizmann Institute of Science, Rehovot 7610001, Israel

[*]Corresponding Authors: jkumar5@unl.edu, amit.finkler@weizmann.ac.il, alaraoui2@unl.edu

**Abstract**

Bond-dependent magnetic interactions, strong spin-orbit coupling, and quantum fluctuations in α−$RuCl_3$ make it an archetypical candidate for exotic Kitaev quantum spin liquid (QSL) states. While electronic transport, neutron scattering, optical and tunneling spectroscopies were used to probe global QSL signatures in large scale α−$RuCl_3$ samples, nanoscale evidence of local QSL behavior via a non-invasive experimental probe is still missing. Here, we use diamond nitrogen-vacancy (NV) spin-noise spectroscopy to probe QSL states in α−$RuCl_3$ nanoflakes with thicknesses ranging from 8 to 150 nm over a temperature range of 4.3–250 K. Our measurements reveal a pronounced decrease in the NV spin-lattice relaxation time ($T_1$) below ∼100 K in proximity to the α−$RuCl_3$ flakes, indicating enhanced GHz-frequency magnetic fluctuations. This temperature scale is consistent with the energy scale associated with Kitaev interactions (∼8 meV) and the emergence of low-energy fractionalized excitations. Below 50 K, thicker (> 90 nm) flakes exhibit a distinct bi-exponential $T_1$ decay, unlike the single-exponential behavior observed for off-flake and thinner (8 nm) flake measurements, suggesting an increasingly heterogeneous magnetic noise environment. These results establish NV spin-noise spectroscopy as a local, non-invasive probe of low-energy spin dynamics in α−$RuCl_3$ nanoflakes and open a route to testing predicted noise signatures of fractionalized excitations in atomically thin QSL candidates.

Quantum spin liquids (QSLs) represent an exotic state of matter that supports fractionalized excitations, such as itinerant Majorana fermions and stationery $\mathbb{Z}_2$ (visons) gauge fluxes[1,2]. The fractionalized quasiparticles obey non-trivial statistics and hold significant potential for realizing noise-resilient qubits for futuristic fault-tolerant quantum computing[3–5]. Among various Kitaev QSLs candidates that have been investigated, the Mott insulator α−$RuCl_3$ has emerged as a promising candidate[6]. In α–$RuCl_3$, a strong spin-orbit coupling supports effective low-energy $J_{eff}$ = 1/2 pseudospins in $Ru^{3+}$ ($4d^5$) electronic state (see Fig. 1a), making parameters favorable for bond-dependent Kitaev interactions[7–9]. Macroscopic magnetic and thermodynamic measurements reveal two closely spaced magnetic phase transitions near 8 K and 14 K, below which spins order antiferromagnetically in a zig-zag configuration in the honeycomb plane[10–12]. In addition to the low-temperature magnetic transitions, structural phase transitions near ∼160 K and ∼60 K have also been reported[13–15].

Beyond conventional magnetic states, the search for fractionalized excitations associated with Kitaev interactions has become a major focus of research in α–$RuCl_3$[1,16,17]. Several theoretical and experimental studies suggested that the conventional Kitaev interactions are significantly stronger in comparison to Heisenberg exchange interactions, with an estimated energy scale of ∼5 to 15 meV, and the signatures of Majorana fermions and $\mathbb{Z}_2$ visons manifesting well above the magnetic ordering temperature[1,18–20]. For instance, inelastic neutron scattering revealed a broad magnetic excitation continuum in α–$RuCl_3$ extending up to ∼20 meV and persisting up to temperatures of approximately 120 K, well above the antiferromagnetic ordering temperature and within the energy scale associated with Kitaev interactions[7,21–23]. Raman scattering is also a powerful experimental probe of Kitaev interactions, with a broad scattering continuum expected as a characteristic signature of spin fractionalization into Majorana fermions[24,25]. Indeed, temperature-dependent Raman spectroscopy of α–$RuCl_3$ showed a broad continuum background developing below ~100 K, alongside Fano resonances indicative of strong coupling between the magnetic continuum and the lattice dynamics, supporting the presence of Kitaev interactions in α–$RuCl_3$[26,27]. However, no single experimental technique, to date, has provided an unambiguous "smoking-gun" signature of fractionalized excitations in QSLs, and the suitable experimental probes that can provide complementary and high sensitivity evidence remains elusive.

Recently, nitrogen-vacancy (NV) centers in diamond have emerged as versatile quantum sensors for temperature, pressure, electric and magnetic fields with very high sensitivity and resolution[28–36]. In the realm of quantum sensing, NV centers offer unmatched advantages over conventional experimental techniques widely used to explore low-dimensional magnetic systems, owing to their broadband frequency response, nanoscale sample-to-sensor proximity, and no requirement for large sample sizes[28,35,37]. While the characteristic energy scale of fractionalized excitations vastly exceeds microwave (MW) resonance frequencies available in NV magnetometry in the low magnetic field regime, NV center-based spin-noise spectroscopy may provide a powerful pathway around this limitation. Specifically, NV spin-lattice relaxation time ($T_1$) based relaxometry[38–43] is sensitive to magnetic noise in the GHz regime and can work as a low-energy spectroscopic extension of Raman and neutron scattering techniques. Considering the NV center's excellent sensitivity to magnetic and electric field noise and drawing inspiration from recent theoretical proposals[44–49] to employ NV $T_1$ relaxometry for exploring QSLs, there is significant interest in using spin-qubit magnetometry to probe QSLs in α–$RuCl_3$.

Here, we probe the quantum noise from α–$RuCl_3$ flakes (thickness of 8 – 150 nm) using NV $T_1$ relaxometry from 4.3 to 250 K under various external magnetic fields. Our measurements capture signatures of magnetic transitions through distinct anomalies in $T_1$. More importantly, the temperature-dependent $T_1$ exhibits a sharp drop below ~100 K, a temperature scale that is broadly consistent with the estimated Kitaev interaction energy scale reported for α–$RuCl_3$[7,18,21,21,27,50]. This scale also coincides with the temperature regime in which broad magnetic continuum have been reported in Raman and neutron scattering measurements. Furthermore, our measurements reveal a thickness-dependent anomaly: thicker (> 90 nm) flakes exhibit a crossover from mono-exponential to bi-exponential $T_1$ decay below 50 K, characterized by distinct fast and slow relaxation decay components. This bifurcation emerges near a temperature comparable to the ~40 K scale at which magnetic entropy release has been reported in thermodynamic and optical studies, suggesting a substantial change in the spin dynamics of thicker α−$RuCl_3$ flakes. These results demonstrate that NV quantum sensors can probe low-energy magnetic dynamics in QSLs and provide local insight into their temperature-dependent fluctuations.

## Results and Discussion

Recently, several theoretical studies proposed using NV spin-noise spectroscopy to diagnose magnetic ground states in magnetic insulators, including magnons in magnetically ordered states, spinons paring phase transition from $U(1)$ to $\mathbb{Z}_2$ in spin liquids, and fractionalized excitations in QSLs[44–47]. Fig. 1b illustrates the experimental concept, in which NV centers, positioned proximate to the α−$RuCl_3$ flake, probe local and time-dependent magnetic-field fluctuations generated by the underlying Ru spins. The magnetically active Ru ions form honeycomb layers within the *a-b* plane, thus providing venue for bond-dependent Kitaev interactions. The crystal structure shown in Fig. 1a was generated using VESTA software[51]. In the spin fractionalized picture, itinerant Majorana fermions are coupled to an emergent $\mathbb{Z}_2$ gauge-flux background, giving rise to characteristic spin dynamics and associated magnetic-field fluctuations[7,23]. These fluctuations contribute to the magnetic noise sensed by the nearby NV center and are probed through measurements of its $T_1$.

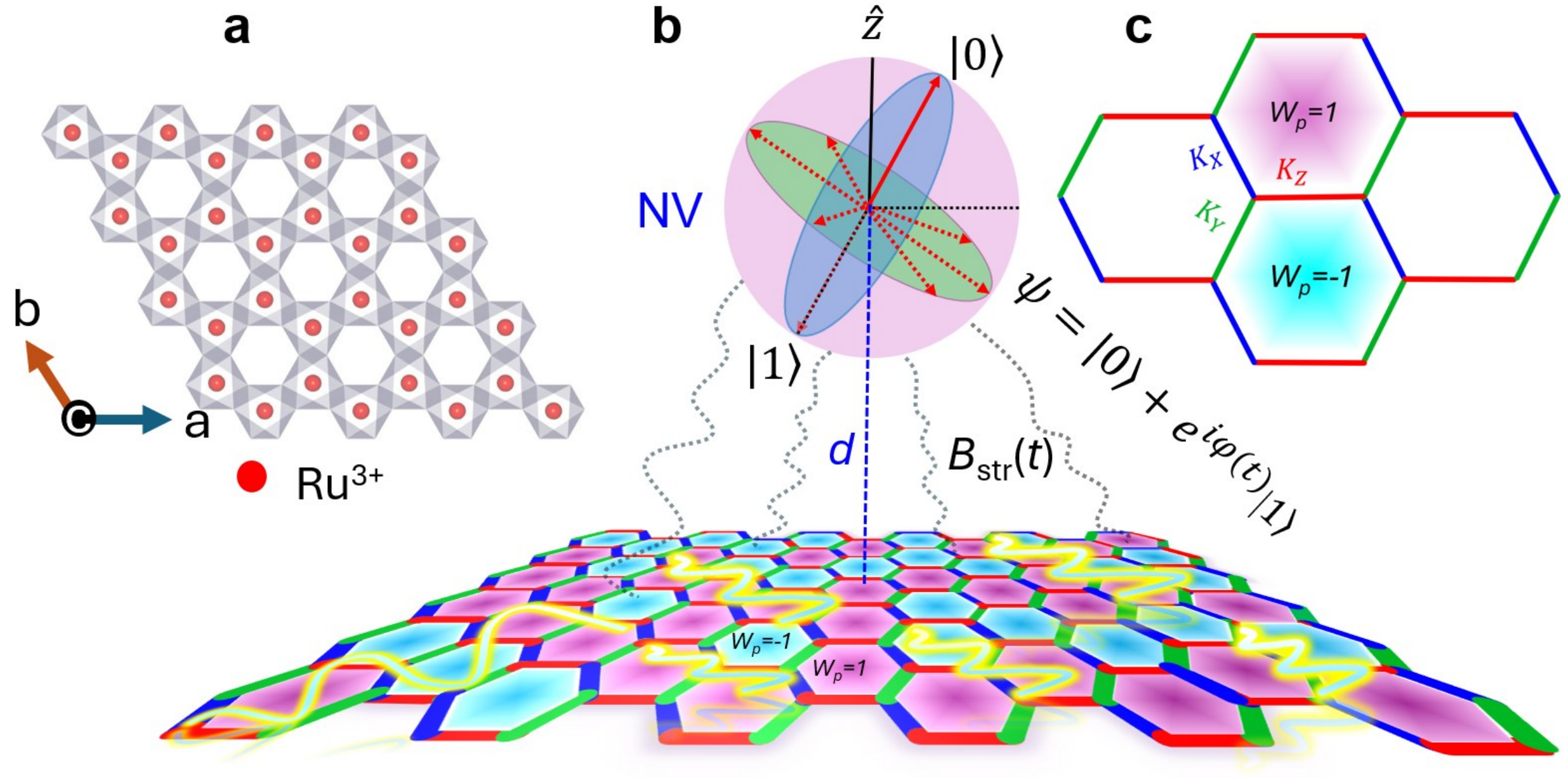

**Fig.1 The concept of NV spin sensing of quantum spin liquid states in α−$RuCl_3$. a** Crystal structure of α−$RuCl_3$ where $Ru^{3+}$ ions form a honeycomb lattice in the *a-b* plane, produced using VESTA software[51]. **b** Schematic of the honeycomb lattice with bond-dependent Ising-type magnetic interactions $K_X$, $K_Y$, and $K_Z$ between Ru electronic spins. $W_p = \pm 1$ denotes the $\mathbb{Z}_2$ gauge-flux sector associated with each hexagonal plaquette as shown in **c**. At finite temperature, itinerant Majorana excitations propagate through the $\mathbb{Z}_2$ gauge-field background. The spin fluctuations generate time-dependent local magnetic stray field $B_{str}(t)$ whose stochastic component generates magnetic noise at the NV standoff d from the α−$RuCl_3$ surface.

To investigate the spin dynamics of α–$RuCl_3$ nanoflakes over the temperature range of 4.3 – 250 K, we first performed $T_1$ relaxometry measurements on the 8 nm thick flake (flake-1) at two magnetic fields (9.59 and 27.86 mT). See Ref. 43 for the synthesis of α-$RuCl_3$ single crystals and Methods for details on flake exfoliation, thickness determination, and transfer onto diamond. The corresponding $T_1$ curves and fits are provided in the Supplementary Information Section S2, Fig. S4, and Fig. S5. Figure 2a shows the temperature dependence $T_1$ measured on and off the α–$RuCl_3$ flake-1 at 9.59 mT, applied along the NV center quantization axis. $T_1$ exhibits a similar temperature dependence for the on-flake and off-flake measurements from 250 K down to ~160 K. In this temperature range, $T_1$ increases upon cooling, with comparable $T_1$ times for both measurement locations. Such behavior is consistent with relaxation dominated primarily by background or surface paramagnetic spin noise in the diamond environment[52]. Below ~160 K, however, the on and off-flake $T_1$ curves begin to diverge. While the off-flake $T_1$ continues to increase upon cooling, the on-flake $T_1$ remains nearly constant between ~160 and ~100 K, within the experimental uncertainty, before exhibiting a pronounced decrease below ~100 K. The off-flake raw $T_1$ curves with fitting are provided in Supplementary Information Section S2 and Fig. S6. The deviation from the high temperature paramagnetic Curie-Weiss behavior in bulk magnetic susceptibility also coincides with this temperature discussed below (Fig. 3b). The onset of this divergence near ~160 K occurs close to the structural transition of α–$RuCl_3$, suggesting that structural change in the flake may contribute to an additional magnetic-noise contribution from the flake[13,14]. At lower temperatures, additional features are observed in the on-flake $T_1$ data, including pronounced dips near ~14 K and ~7 K. These temperatures are associated with the antiferromagnetic phase transitions reported in α-$RuCl_3$ [10–12], where the magnetic noise is enhanced[53].

A comparison of the NV $T_1$ (on and off-flake) in Fig. 2a shows a sharp decrease of $T_1$ below ~100 K in the on-flake condition, indicating an increase in the magnetic noise originating from α-$RuCl_3$. Interestingly, this temperature range corroborates with the previous Raman and Neutron spectroscopy observations in α–$RuCl_3$[7,22,27]. For instance, temperature-dependent Raman measurements have reported unusual magnetic scattering[27], characterized by a broad continuum persisting over a wide temperature range up to 100 K, above the magnetic ordering. It coincides with the energy scale of Kitaev exchange interactions $J_K$ ~8 meV, giving $J_K$ /$k_B$ ~93 K, where $k_B$ is the Boltzmann constant. An asymmetric Fano the line shape of phonon modes also shows anomalous temperature behavior around the same temperature range[27]. In general, a Fano-line usually arises from the interaction between discrete phonon modes and electronic states, signifying metallicity in phase-separated systems[54–56]. The magnetic contribution to the heat capacity of α–$RuCl_3$ also shows linear temperature dependence between 100 K to 50 K, indicative of a metal like density of state characteristic of itinerant Majorana fermions emerging < 100 K[23]. As the dipolar coupling is temperature-independent, the temperature dependence of the NV center's $T_1$ is governed by the variation in the noise spectral density of α–$RuCl_3$ with temperature. Given the clear signature of an increased spectral weight associated with the magnetic scattering continuum below 100 K observed in Raman spectroscopy and neutron scattering, we can attribute our NV spin-noise spectroscopy results to the GHz-frequency extension of the magnetic continuum observed in Raman and neutron scattering studies.

Thus, the decrease in $T_1$ below 100 K indicates that the magnetic fluctuations in the GHz frequency range become stronger as the temperature is lowered. This also suggests that the change in the low-energy magnetic fluctuations occurs near the characteristic energy scale of the Kitaev interactions. This behavior may be related to the evolution of low-energy magnetic excitations in the intermediate-temperature fractionalized regime[57,58], where thermally excited $Z_2$ fluxes modify the itinerant Majorana spectrum and generate finite low-energy spectral weight.

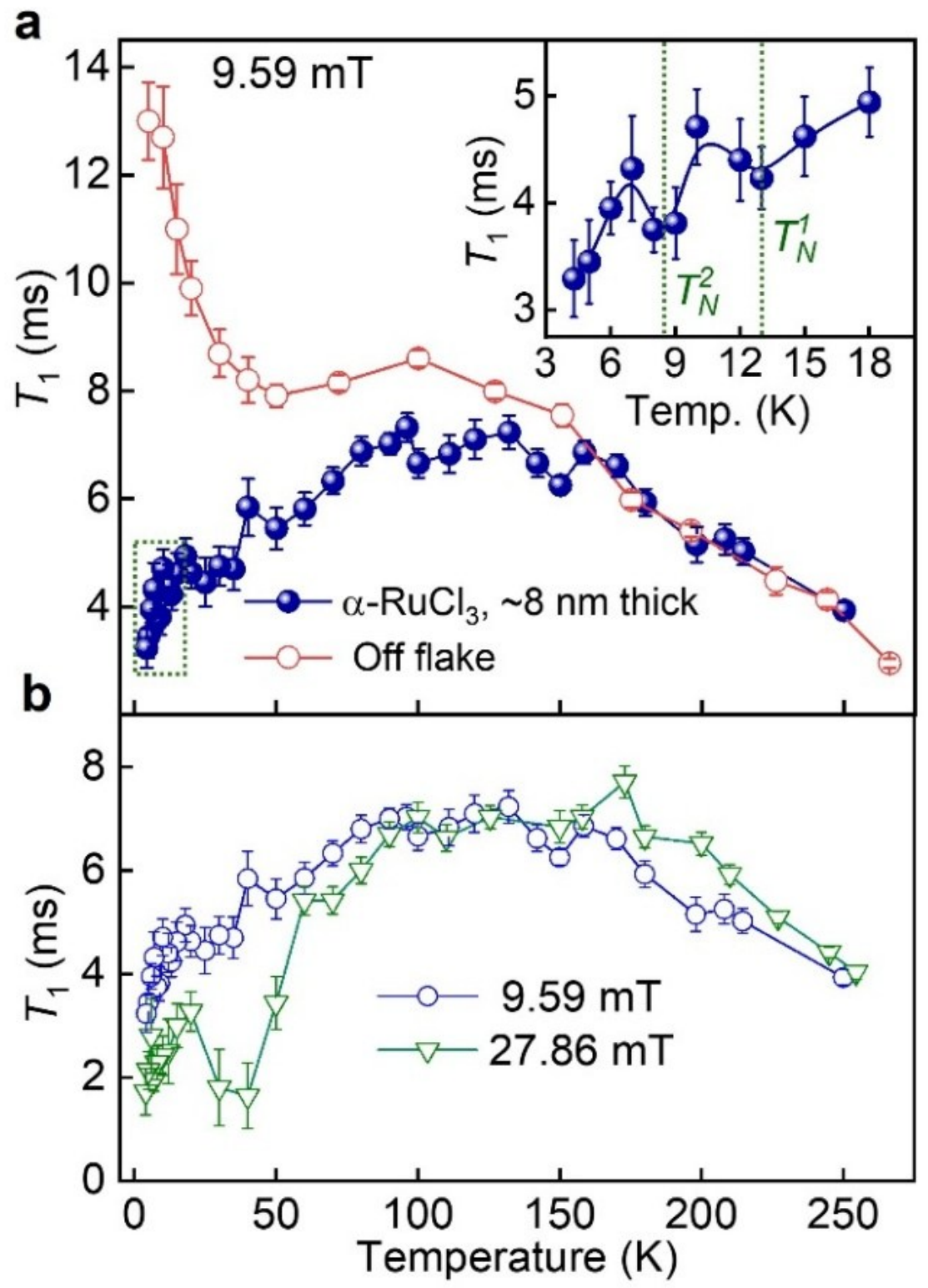

**Fig. 2 NV spin-noise spectroscopy of α–RuCl$_3$ nanoflake**. **a** Temperature-dependent NV $T_1$ measured off and on flake-1 (~8 nm thick) α–RuCl$_3$ flake at a magnetic field of 9.59 mT. The inset shows a zoomed view of the on-flake $T_1$ values, highlighting the two antiferromagnetic phase transitions in α–RuCl$_3$. **b** A comparison of the temperature-dependent NV $T_1$ on flake-1 at magnetic fields of 9.59 mT (open circles) and 27.86 mT (open triangles).

To further explore the frequency dependence of the magnetic noise, we varied the external magnetic field $B$ and performed $T_1$ relaxometry with the field applied along the NV quantization axis. The magnetic field produces a Zeeman splitting of the NV spin levels, shifting the transition frequencies linearly with magnetic field. In our measurements, we used the lower NV frequency ($f_-$) transition for $T_1$ relaxometry. Thus, increasing the external magnetic field shifts the $f_-$ transition to lower frequencies, allowing us to probe the magnetic noise at different frequencies[59].

Figure 2b shows temperature dependent NV $T_1$ measured on flake-1 for two applied fields, 9.59 and 27.86 mT. At high temperatures (160 – 250 K), $T_1$ is longer at the higher field than at the lower field, consistent with our previous measurements of the field dependence of NV $T_1$ near α–RuCl$_3$ in the paramagnetic regime[43]. Below the structural transition at ~160 K, the relaxation becomes nearly field independent over the 160 – 100 K temperature range. Interestingly, below ~100 K, the $T_1$ curves bifurcate, with $T_1$ decreasing more rapidly at the higher field upon cooling and remaining below the values measured at the lower field. Since the higher field probes the $f_-$ transition at a lower frequency, this behavior suggests that the magnetic noise in α–RuCl$_3$ acquires enhanced spectral weight at lower frequencies at low temperatures. At the higher field, $T_1$ exhibits an upturn (a dip-like) below ~40 K, reaching a maximum near 20 K before decreasing again upon further cooling. This non-monotonic behavior may reflect a modification of the magnetic noise spectrum as the system approaches the long-range antiferromagnetic transition[53].

We next examined the thickness dependence of the magnetic noise in α−$RuCl_3$ flakes: flake-2 (thickness of 95 ± 1 nm) and flake-3 (thickness of 150 ± 1 nm). Figure 3 summarizes the $T_1$ measurements (photoluminescence, PL, vs. time) of flake-2 (thickness ~95 nm), its $T_1$ values are lower than those of flake-1 (thickness of ~8 nm), consistent with a larger magnetic-noise contribution from the thicker flake in the paramagnetic regime[43]. The $T_1$ data of flake-2 and corresponding single and double exponentials decay fits are provided in the Supplementary Information Fig. S7. A comparison of the temperature dependence of $T_1$ for flakes of different thicknesses is given in the Supplementary Information Section S2 and Fig. S8. Figure 3a shows the evolution of the NV $T_1$ from a single-exponential decay at higher temperatures to a two-component decay below ~50 K. Figure 3b compares the $T_1$ curves (normalized PL vs. time) at 10 K measured at 9.59 mT, for the off-flake region, flake-1, flake-2, and flake-3, highlighting the distinct behavior of the thicker flakes. Above 50 K, $T_1$ is obtained from single-exponential fits, the temperature dependence of the extracted $T_1$ components is shown in Fig. 3c, which follows the same trend as the flake-1, i.e., rising until ~160 K and plateau in the window 100 – 160 K, and dropping below ~100 K.

The fast component decreases continuously upon cooling, whereas the slow component increases toward the antiferromagnetic ordering temperature before decreasing at lower temperatures. This trend is clearer in Fig. S8 (Supplementary Information) for the linear-log plot of $T_1$ vs temperature). The transition to two-component relaxation is accompanied by a sharp decrease in the stretching parameter η below ~100 K, from ~0.8 to ~0.37 at 50 K (Fig. 3d). This indicates a progressively broader distribution of relaxation rates, which eventually cannot be adequately captured by a single stretched-exponential decay.

The origin of the bi-exponential $T_1$ decay emerging below ~50 K (see Methods), particularly in the thicker (> 90 nm) flakes, is not yet clear. Bi-exponential relaxation in ensemble NV measurements can arise from a heterogeneous distribution of relaxation rates among individual NV centers, while cross-relaxation between nearby NVs or surface-related NVs can also contribute[60–62]. Given the implantation dose of $10^{13}$ $cm^{-2}$ of our NVs,[42] corresponding to an average NV–NV separation of ~30 nm, the cross-NV relaxation is unlikely to be the dominant mechanism. The low-temperature onset of the bi-exponential behavior also makes a charge-state-related contribution less likely. We therefore attribute the emergence of two resolvable relaxation components to an increasingly heterogeneous magnetic-noise environment in the thicker α−$RuCl_3$ flakes. This interpretation is consistent with the progressively decreasing stretching exponent upon cooling, indicating a broadening distribution of relaxation rates that eventually becomes resolvable as a bi-exponential decay. Thus, the bi-exponential response provides evidence for heterogeneous low-temperature magnetic dynamics. Interestingly, while the exact microscopic mechanism requires further study, specific heat measurements of α−$RuCl_3$ crystals reveal a broad maximum in the range $T_N < T < 50$ K, attributed to the localized Majorana fermion excitations[23], that closely matches the inception temperature of our bi-exponential feature. Furthermore, the emergence of an additional relaxation component below ~50 K occurs in a temperature regime where low-energy magnetic correlations become increasingly important, distinct from the higher exchange-energy scale $|K|$~ 8 meV reported in theoretical and experimental studies[23,63–66].

For instance, bifurcation of $T_1$ below ~50 K is consistent with the Kitaev crossover scale temperature $T_H \sim 0.375|K|$ ($K$ is the Kitaev energy scale), below which optical spectroscopy reveals a marked redistribution of spectral weight associated with the development of bond-dependent nearest-neighbor spin correlations[63]. This scale also marks the thermal fractionalization of quantum spins into itinerant Majorana fermions, signaling entry into a Kitaev-correlation-dominated regime[65]. Theoretical calculations for the Kitaev model predict an enhancement of the nuclear magnetic resonance (NMR) spin-lattice relaxation rate $1/T_1$ below $T_H$ as the Ru electronic spins become progressively correlated[64], and the additional NV relaxation component observed here may reflect the same underlying growth of electronic spin fluctuations in this temperature regime. This low-temperature feature may therefore reflect a change in the magnetic fluctuation spectrum, although its microscopic origin remains unclear. Systematic layer-by-layer, thickness-dependent measurements will be valuable for elucidating the nature of this relaxation channel below 50 K.

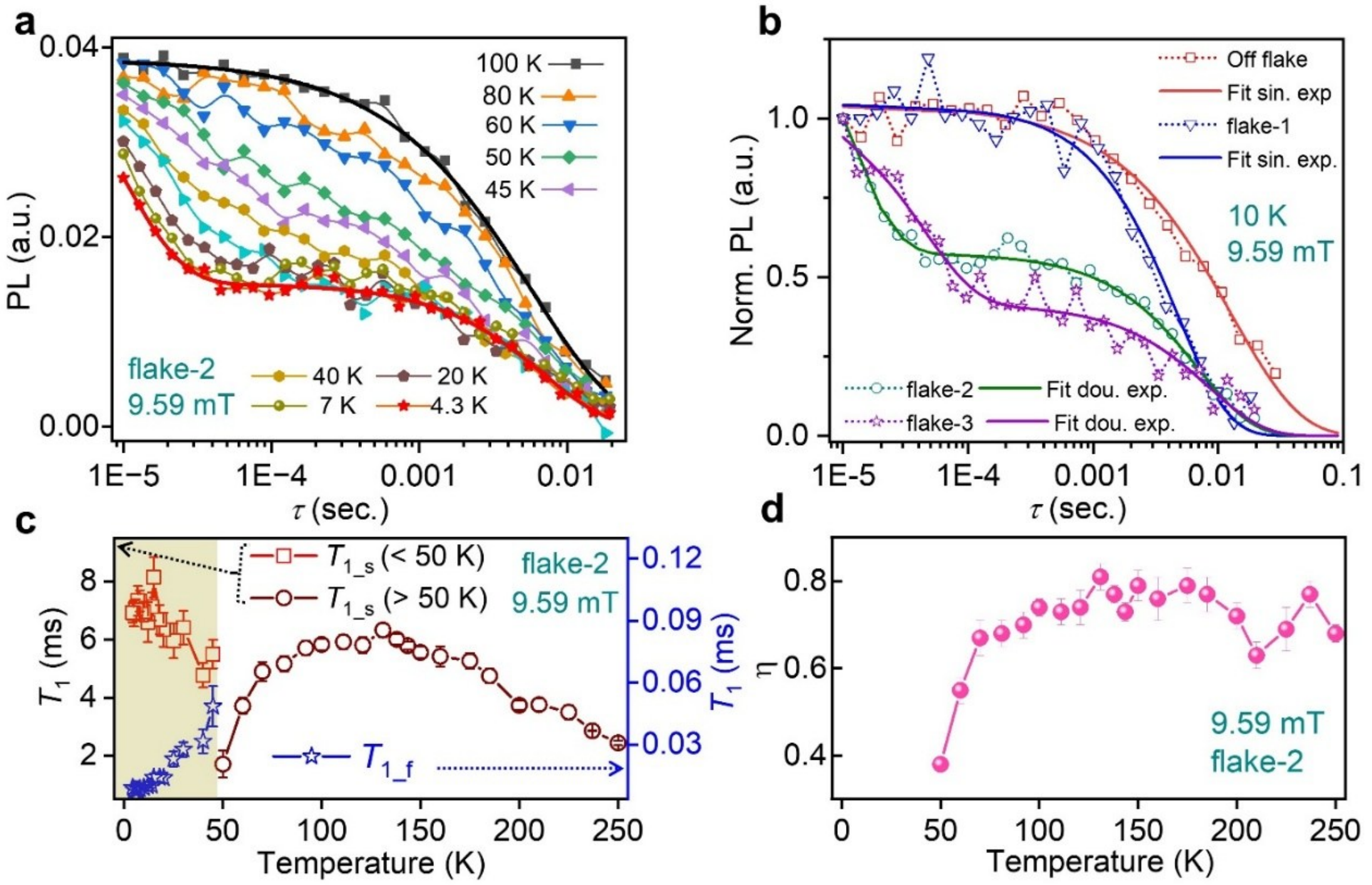

**Fig. 3 Effect of α–$RuCl_3$ flakes' thickness on NV-$T_1$.** **a** NV $T_1$ curves (PL vs time) at different temperatures, showing the evolution from single-exponential relaxation at high temperatures to double-exponential relaxation at low temperatures. **b** Comparison of the $T_1$ curves (normalized PL vs time) at 10 K and at 9.59 mT measured off the flake and on flake-1 (~8 nm), flake-2 (~95 nm), and flake-3 (~150 nm). The off-flake and thin flake 1 data are fitted using a single-exponential decay function (equation (5) in Methods, solid lines), whereas the thicker flake-2 and flake-3 are fitted using a double-exponential decay function (equation (6) in Methods, solid lines). **c** Temperature-dependent $T_1$ measured on flake-2. The fast and slow decays are extracted from the double-exponential fits, see Methods Supplementary Information Section S2 and Fig. S7. **d** Temperature dependence of the stretching exponent $\eta$, obtained from stretched-exponential fits to the on-flake $T_1$ data.

To further examine the correspondence between the local magnetic fluctuations probed by NV-$T_1$ relaxometry and the bulk magnetic response, we compare the $T_1$ data obtained from the thicker flake-3 (Fig. 4a) under 27.66 mT filed with bulk magnetic susceptibility measured by Superconducting Quantum Interference Device (SQUID) magnetometry (Fig. 4b). As shown in Fig. 4a, the temperature dependence of $T_1$ exhibits a distinct change in behavior around ~160 K. Interestingly, the inverse static susceptibility, after subtraction of the Curie–Weiss contribution, also deviates from the paramagnetic behavior in the same temperature range, indicating the development of short-range magnetic correlations[23]. As highlighted in the inset of Fig. 4b, the close correspondence between these characteristic temperature scales suggests that the change in local magnetic-field fluctuations detected by the NV center is associated with the evolution of magnetic correlations in α−$RuCl_3$. Since NV-$T_1$ relaxometry probes magnetic fluctuations near the NV transition frequency, this correspondence between local dynamical relaxation and bulk magnetic susceptibility supports an intrinsic magnetic origin of the observed $T_1$ anomaly. At low temperatures, similar to flake-1 and flake-2, flake 3 exhibits a pronounced decrease in $T_1$ below ~100 K[23], consistent with enhanced low-energy magnetic fluctuations in this regime[6]. At still lower temperatures, flake-3, like flake-2, develops a bi-exponential relaxation below ~50 K. The raw $T_1$ data with corresponding fast and slow $T_1$ components at several temperatures shown in Fig. S11 in the Supplementary Information. The stretching exponent η for flake-3 also decreases markedly below ~100 K shown in the inset of Fig 3a, indicating an increasingly broad distribution of relaxation rates and providing a continuous evolution toward the bi-exponential relaxation observed below 50 K. Consistent with the data collected at 27.66 mT, a similar temperature-dependent behavior of $T_1$ is also observed at a lower field of 9.59 mT, as shown in the Supplementary Information (Fig. S9 and Fig. S10).

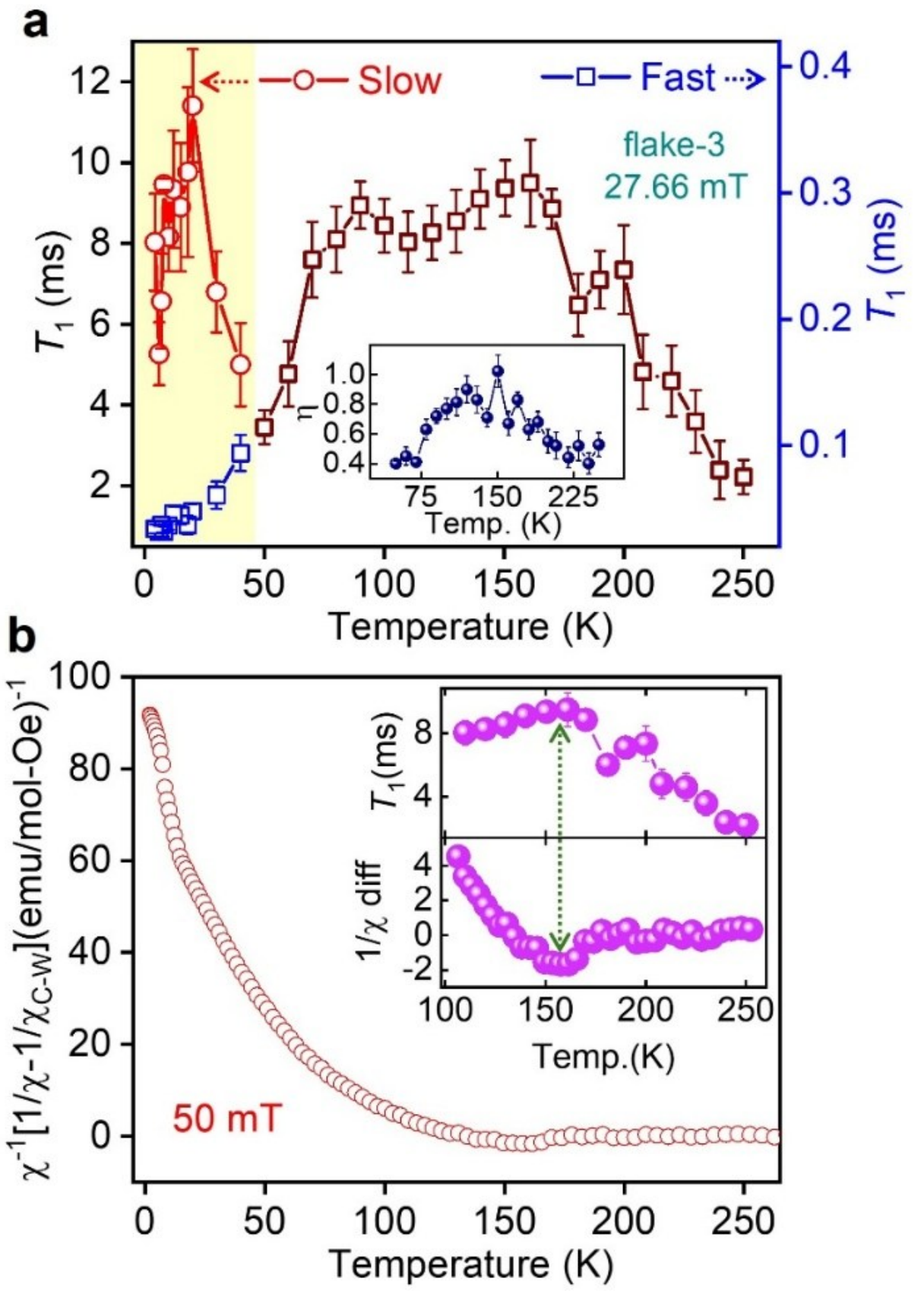

**Fig. 4. Comparison between NV and SQUID measurements on thick α−$RuCl_3$ flakes. a** Temperature dependence of the NV $T_1$ beneath the thickest α-$RuCl_3$ flake (flake-3). Inset of (**a**) shows $\eta$ vs temperature, highlighting a maximum ~150 K, like $T_{1_s}$. **b** The difference of inverse magnetic susceptibility obtained by subtracting the Curie–Weiss fit from the measured data from SQUID magnetometer. The inset of (**b**) highlights that both ($T_1$) and the susceptibility difference exhibit a transition at approximately the same temperature (~160 K).

In summary, we investigated the spin dynamics of nanometer-thick $\alpha-RuCl_3$ nanoflakes using NV $T_1$ relaxometry. The NV relaxation exhibits a clear anomaly near ~160 K, in close correspondence with our bulk magnetization measurements, followed by a broad plateau between ~ 100 and 160 K, and a pronounced decrease in $T_1$ below ~100 K. This behavior indicates an enhancement of GHz frequency magnetic noise on a temperature scale comparable to the Kitaev exchange-energy scale, $|K|$~8 meV. The enhancement of the magnetic noise spectral density as measured by the NV centers, is consistent with dynamic features associated with itinerant Majorana fermions and thermally excited $\mathbb{Z}_2$ gauge fluxes[6,7,23]. The emergence of bi-exponential NV $T_1$ relaxation below 50 K in thicker (> 90 nm) flakes indicates an additional change in the local magnetic dynamics and closely relates with the spin dynamics regime change calculated theoretically and observed in thermodynamic and optical measurements.

Overall, our results demonstrate that NV spin relaxometry provides a local, frequency-selective probe of low-energy magnetic fluctuations in $\alpha-RuCl_3$, complementary to Raman and neutron scattering spectroscopies [6]. It offers a sensitive approach for exploring low-frequency spin dynamics in other candidate QSL materials[1,2,67,68]. Future layer-by-layer noise profile qubit measurements at higher magnetic field may provide more information about the complex dynamics in QSLs.

## Methods

**Sample preparation and characterization**. To investigate $\alpha-RuCl_3$ at the nanoscale, we used shallow NV centers created by nitrogen implantation ~6 nm beneath the surface of a (100)-oriented diamond[41]. $\alpha$-$RuCl_3$ nanoflakes were mechanically exfoliated from bulk crystals using Scotch tape and afterwards transferred onto the diamond chip via a silicone gel film using a micromanipulator transfer stage[36,42]. To achieve better thermal contact with the cold head of the closed-cycle refrigerator, the diamond was mounted on a sapphire sample platform patterned with gold transmission lines for microwave delivery and NV-center spin state manipulation, as described in our previous work[42]. Details of the $\alpha-RuCl_3$ crystal growth are provided in Ref. 43. NV fluorescence measurements were performed using a home-built cryogenic NV microscopy setup, as described previously[42]. The flakes' thickness was determined by atomic force microscopy (AFM). We focused on three representative flakes with thicknesses of 8 ± 0.5 nm for flake-1, 95 ± 1 nm for flake-2, and 150 ± 1 nm for flake-3. The corresponding AFM height profiles and optical images are provided in Supplementary Information Section S1 (Fig. S1, Fig. S2, and Fig. S3).

**NV Magnetometry**. Consider the NV center as an effective two-level system from the spin-1 ground state with resonance frequency $\omega = (D \pm \gamma B)$ , where $D$ = 2.87 GHz is the zero-field splitting, $\gamma$ is the NV gyromagnetic ratio, and $B$ is the external magnetic field applied along the NV quantization axis. The interaction between the NV spin and the time-dependent magnetic stray field $B_{str}(t)$ generated by spin fluctuations in the underlying two-dimensional (2D) honeycomb lattice, is described by the Hamiltonian:[30]

$$H_{int}(t) = g\mu_B \boldsymbol{S} \cdot \boldsymbol{B}_{str}(t), \quad (1)$$

where $g$ is the electronic factor, $\mu_B$ is the Bohr magneton and $\boldsymbol{S} = (S_x, S_y, S_z)$ represents the dimensionless spin-1 operators of the NV electron spin. The key information of magnetic noise is encoded in the associated magnetic noise power spectral density (PSD). The $T_1$ relaxation time of NV center is more sensitive to the transverse noise component perpendicular to the quantization axis [41–43]. At equilibrium temperature, the PSD of a spin bath can be expressed as[38,48,69]:

$$S_{B\perp}(\omega) = \int_{-\infty}^{+\infty} dt e^{i\omega t} \langle B_{str\perp}(t) B_{str.\perp}(0) \rangle, \tag{2}$$

In corollary, within the Fermi Golden rule for the transition rate $\Gamma = 1/T_1$ of the NV spin population in spin states $|0\rangle$ and $|\pm 1\rangle$ in the presence of PSD can be expressed as[38,49]:

$$\frac{1}{T_1} = \left(\frac{1}{T_1}\right)_i + \int \gamma^2 \langle B_{str}^2 \rangle \, S(\omega, T, E_a) F(\omega) \, d\omega \,, \tag{3}$$

where $\left(\frac{1}{T_1}\right)_i$ refers to the intrinsic relaxation mechanisms, $\sqrt{\langle B_{str}^2 \rangle}$ is the root-mean-square value of the effective magnetic field at the NV position, and $E_a$ is the energy of the magnetic anisotropy of the flake. $S(\omega)$ is the temperature dependent spectral density of the spin-noise, and $F(\omega)$ is a filter function[60]:

$$F_1(\omega) = \sum_{i=\mp} \frac{4\pi/T_2^*}{(2\pi/T_2^*)^2 + (\omega - \omega_i)^2} \,, \tag{4}$$

where $T_2^*$ is the NV dephasing time and $\omega_i$ is the NV resonance frequency at zero magnetic field (~2.87 GHz).

NV spin-noise spectroscopy measurements were performed by first optically polarizing the NV into the $m_s = |0\rangle$ spins state using a 532 nm green laser (power of ~38 mW, beam diameter of 23 μm). A MW π pulse, with the duration determined from Rabi oscillations, was then applied to transfer the NV population from $m_s = |0\rangle$ state to one of the $m_s = |\pm 1\rangle$ states. Following a variable waiting dark time $\tau$, a laser readout pulse was applied to measure the remaining $m_s = |0\rangle$ spin population by collecting the red (650 – 800 nm) fluorescence. The measured fluorescence was then subtracted from the fluorescence obtained using an identical pulse sequence without the MW π pulse[42,52]. The $T_1$ time is calculated by fitting the fluorescence vs $\tau$ curve to a stretched exponential decay function:

$$P(\tau) = \tau_0 e^{\left(-\frac{\tau}{T_1}\right)^{\eta}} + C, \tag{5}$$

where $P(\tau)$ is the NV fluorescence signal, $\tau_0$ is a pre-factor, $T_1$ is spin-lattice relaxation time of NV, $\eta$ is a stretching parameter, and $C$ is offset parameter.

Below 50 K, the fast and slow components of $T_1$ on thicker flakes (flake-2 and flake-3) were extracted from double-exponential fits as[60–62]:

$$P(\tau) = \tau_{0F} e^{(-\tau/T_{1_f})} + \tau_{0S} e^{(-\tau/T_{1_s})}, \tag{6}$$

where $P(\tau)$ is the NV fluorescence intensity measured at different times $\tau$. $\tau_{0F}$ and $\tau_{0S}$ are the prefactors related to the fast $T_{1_f}$ and $T_{1_s}$ decay times, retrospectively.

## Data Availability Statement

The data that support the findings of this study are available from the corresponding author upon reasonable request.

## Acknowledgements

A.L. and X.H. acknowledge the National Science Foundation (NSF) Award 2521415 and the University of Nebraska-Lincoln (UNL) Grand Challenges catalyst award entitled "Quantum Approaches addressing Global Threats". S.-H.L., X.H., and A.L. acknowledge additional support from NSF EPSCoR RII Track-1: Emergent Quantum Materials and Technologies Award 2044049.

The research done at UNL was performed in part in the Nebraska Nanoscale Facility: National Nanotechnology Coordinated Infrastructure and the Nebraska Center for Materials and Nanoscience (and/or NERCF), supported by NSF Award 2025298. M.H. acknowledges funding support from the M. and Harry B. Helmsley Charitable Trust, grant no. 2018PG-ISL006. A.F. is the incumbent of the Elaine Blond Career Development Chair in Perpetuity and acknowledges supported by the Minerva Stiftung with funding from the Federal German Ministry for Education and Research. This research is made possible in part by the historic generosity of the Harold Perlman Family.

## Conflict of Interest

The authors declare no conflict of interest.

## Author Contributions

J.K. A.F and A.L. conceived the idea. J.K. and A.L designed the experiments and supervised the project. J.K. performed the NV measurements with assistance from S.I., B.H., and R.T. H.S. performed additional optical and AFM measurements. T.L. and X.H. exfoliated the $\alpha-RuCl_3$ flakes and performed optical and AFM imaging. S.I. and S.-H. L made the MW loops. A.S., A.K.P. and M.H. synthesized the $\alpha-RuCl_3$ crystals. J.K. and A.L. wrote the manuscript with contributions and feedback from all authors.

## Supplementary Information

## Supplementary Note S1. Optical and topography imaging of $\alpha$–$RuCl_3$ flakes

Figure S1a and S1b show optical images of the exfoliated $\alpha$–$RuCl_3$ flakes transferred onto polydimethylsiloxane (PDMS) film and diamond substrates respectively. To determine the thickness of the thinnest flake (flake-1), investigated using the nitrogen-vacancy (NV) microscope in the main text, we performed atomic force microscopy (AFM) measurements. The corresponding AFM image, acquired after transferring the flake onto the diamond substrate, is shown in Figure S1c, and the corresponding height profile is presented in Figure S14, indicating an approximate flake thickness of 8 ± 0.5 nm.

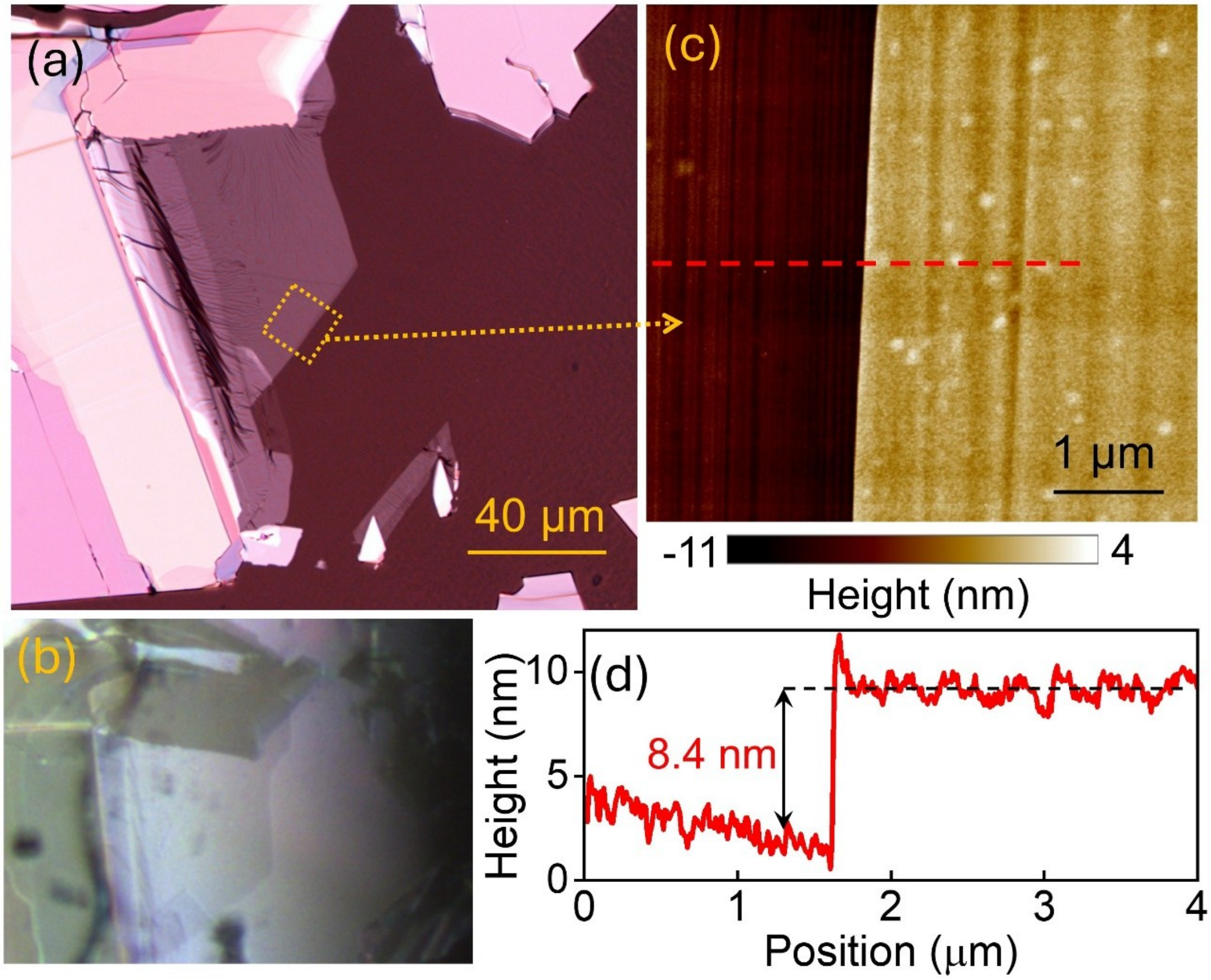

**Fig. S1: Optical microscopy and AFM imaging of $\alpha$–$RuCl_3$ flake-1 investigated in the main text. (a)** Optical image of the flake exfoliated on PDMS. **(b)** Optical image of the flake on the diamond surface. **(c)** AFM image of a region (dashed square) of the flake in **(a)** transferred onto the diamond substrate. **(d)** Height profile along the cross-sectional line marked by the red dashed line in the AFM image in **(c)**, indicating a flake-1 thickness of 8 ± 0.5 nm.

The optical and AFM images of the other two flakes, flake-2 and flake-3, investigated by NV magnetometry during this study are shown in Figures S2 and S3, respectively. The corresponding AFM height profiles indicate approximate thicknesses of 95 ± 1 nm for flake-2 and 150 ± 1 nm for

flake-3. The dots seen in the AFM images come from the contamination of the flakes prior to AFM and after NV measurements due to the slight VGE varnish residue[42].

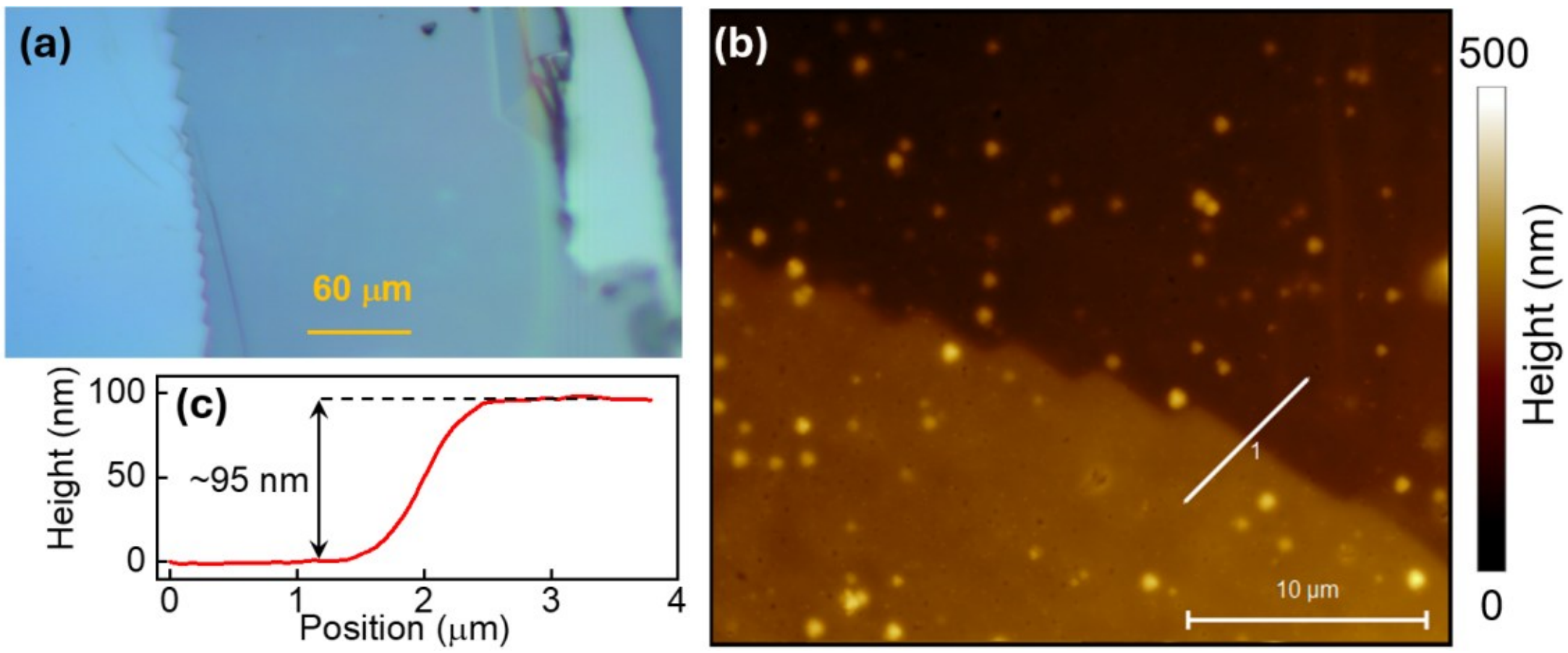

**Fig. S2: Optical microscopy and AFM images of α−$RuCl_3$ flake-2 investigated in the main text. (a)** Optical image of the α-$RuCl_3$ flake transferred into the diamond substrate. **(b)** AFM image of a selected region of the same flake in **(a)**. **(c)** Height profile along the cross-sectional line marked by the white line in the AFM image in **(b)**, indicating a flake-2 thickness of 95 ± 1 nm.

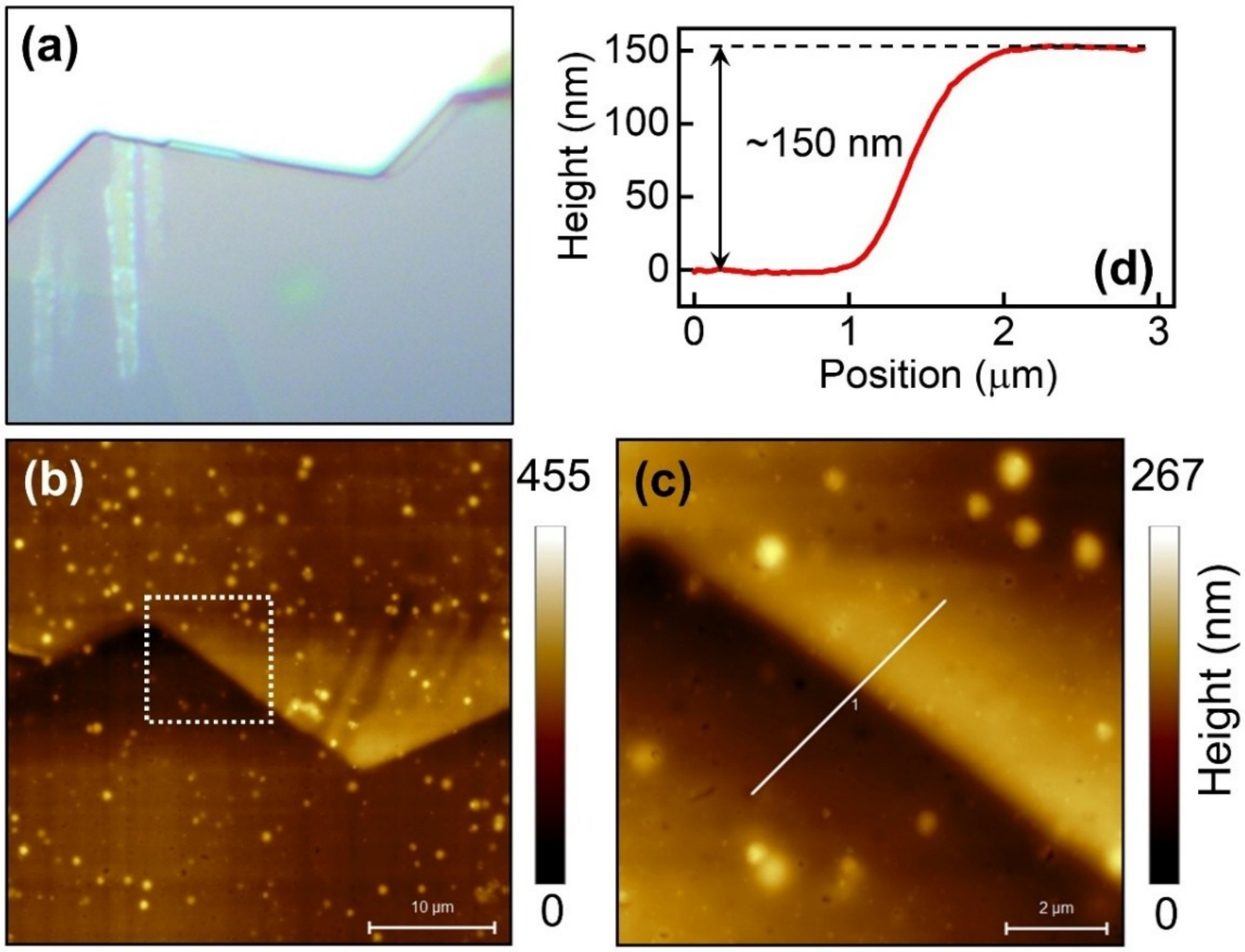

**Fig. S3: AFM and optical microscopy and of α–$RuCl_3$ flake-3 investigated in the main text. (a)** Optical image of part of flake-3 transferred on the diamond substrate. **(b)** AFM image of the flake in **(a)**. **(c)** Zoomed AFM image of the flake-3 in **(b)**, marked with a dashed square. **(d)** Height profile along the cross-sectional line marked by the white line in the AFM image in **(c)**, indicating a flake-3 thickness of 150 ± 1 nm.

## Supplementary Note S2. NV spin lattice relaxation time ($T_1$) on and off $\alpha$–$RuCl_3$ flakes

**$T_1$ raw data under flake-1 at 9.59 mT:** Figure S4 shows the measured $T_1$ curves acquired beneath flake-1 at temperatures ranging from 4.3 to 250 K at an external magnetic field of 9.59 mT. The measured $T_1$ curves were fitted using the stretched-exponential decay function described by equation (5) in the main text, and the extracted $T_1$ values are presented in Figure 2 of the main text.

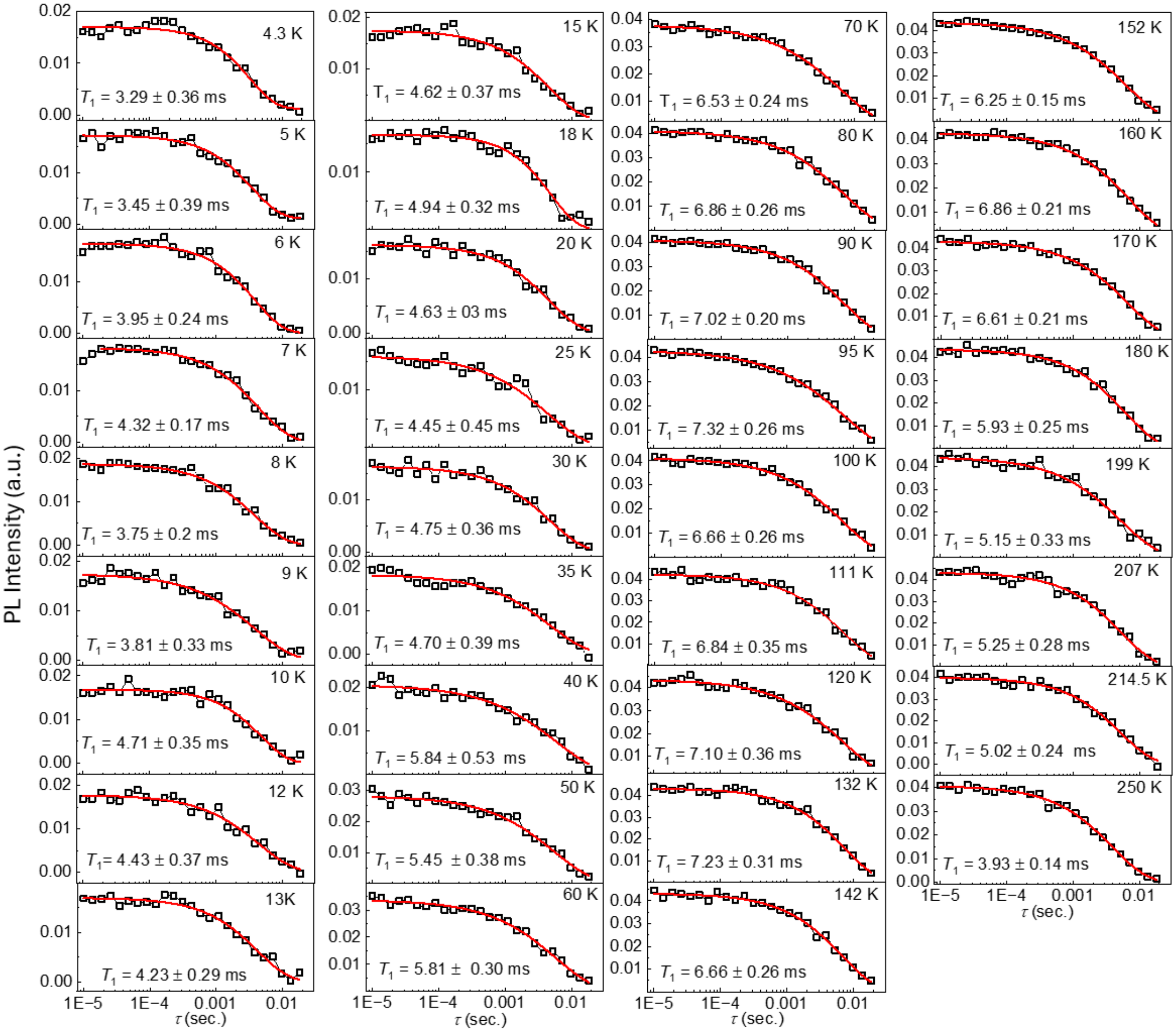

**Fig. S4:** $T_1$ measurements under flake-1 at different temperatures. Open squares symbols are the measured data, and solid red lines are the fits according to equation **(5)** in the main text.

**$T_1$ data under flake-1 at 27.86 mT:** Figure S5 shows the measured $T_1$ curves acquired beneath flake-1 at temperatures ranging from 4.3 to 250 K at an external magnetic field of 27.86 mT. The measured $T_1$ curves were fitted using the stretched-exponential decay function described by equation (5) in the main text, and the extracted $T_1$ values are presented in Figure 2 of the main text.

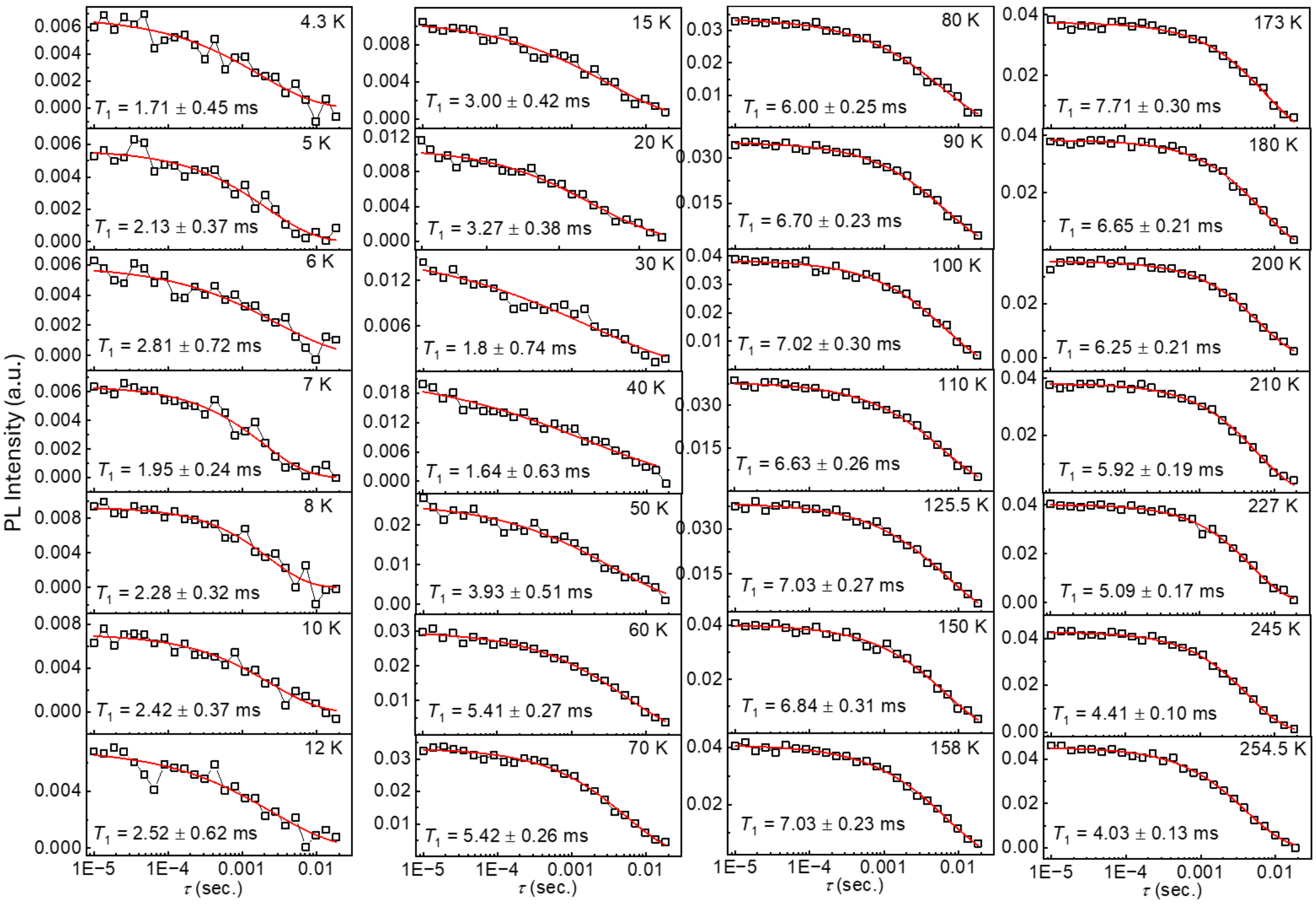

**Figure S5:** Measured (open squares) $T_1$ acquired beneath flake-1 at different temperatures at 27.86 mT. The solid red lines represent fits to the measured data using the stretched-exponential decay function described by equation **(5)** in the main text.

**$T_1$ control data off α−$RuCl_3$ at 9.59 mT:** We performed control measurements of $T_1$ at different temperatures (5 – 266 K) at a location sufficiently far from the α−$RuCl_3$ flakes (see Figure S6). During these measurements, the temperature remained stable within a range of 50–100 mK. The corresponding $T_1$ values were extracted by fitting the measured relaxation curves to a stretched-exponential decay function, as described by equation (5) in the main text. The resulting temperature-dependent $T_1$ data are presented in Figure 2 and discussed in the main text.

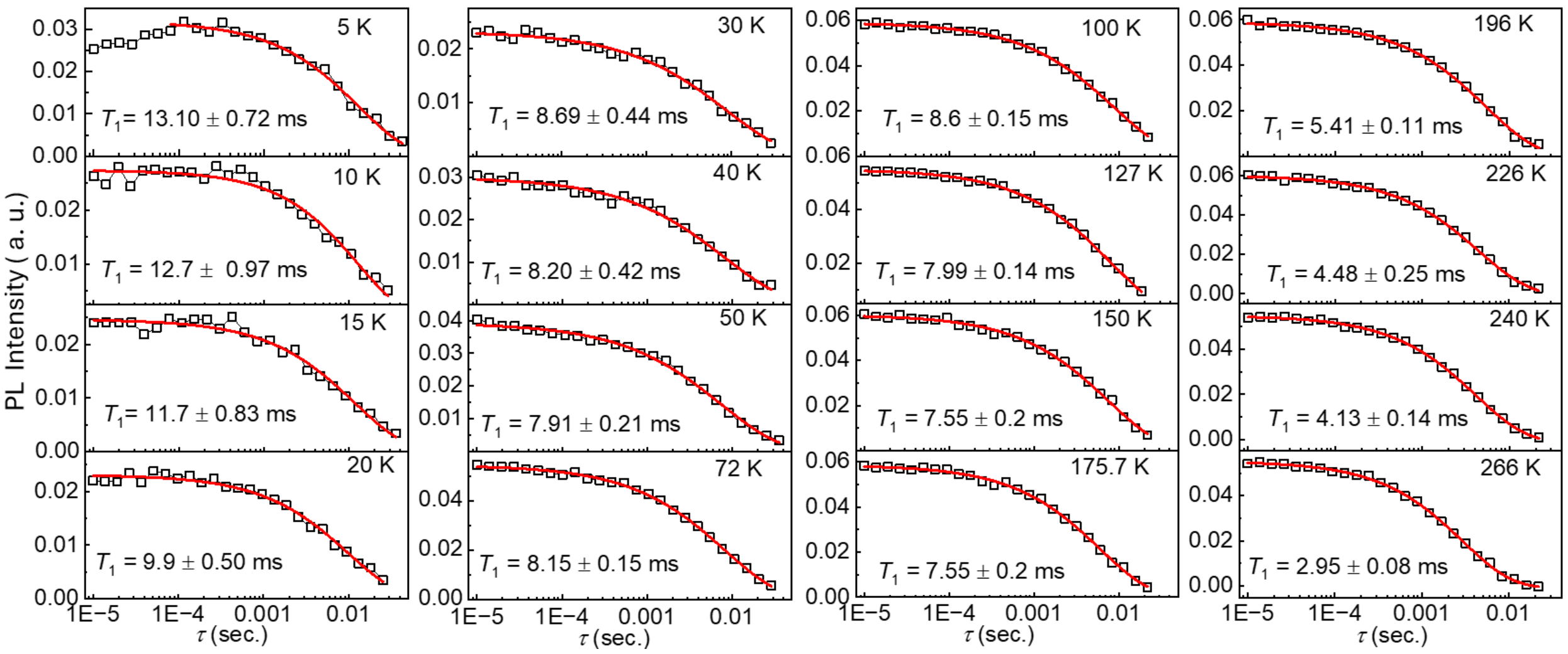

**Figure S6:** Measured (open squares) $T_1$ far away from the α−$RuCl_3$ flakes at 9.59 mT. The solid red lines represent fits to the experimental data using the stretched-exponential decay function described by equation **(5)** in the main text.

***$T_1$* data under α−$RuCl_3$ flake-2 at 9.59 mT:** We measured the temperature dependence of the NV $T_1$ of flake-2 (~95 nm) over a temperature range of 4.3 – 250 K at 9.59 mT. The measured $T_1$ data were fitted using a double-exponential decay model in the temperature range of 4.3 – 45 K, while a single-exponential decay model was used at higher temperatures, as described in the main text. The extracted relaxation times and corresponding fitting results are presented in Figure 3 and discussed in the main text.

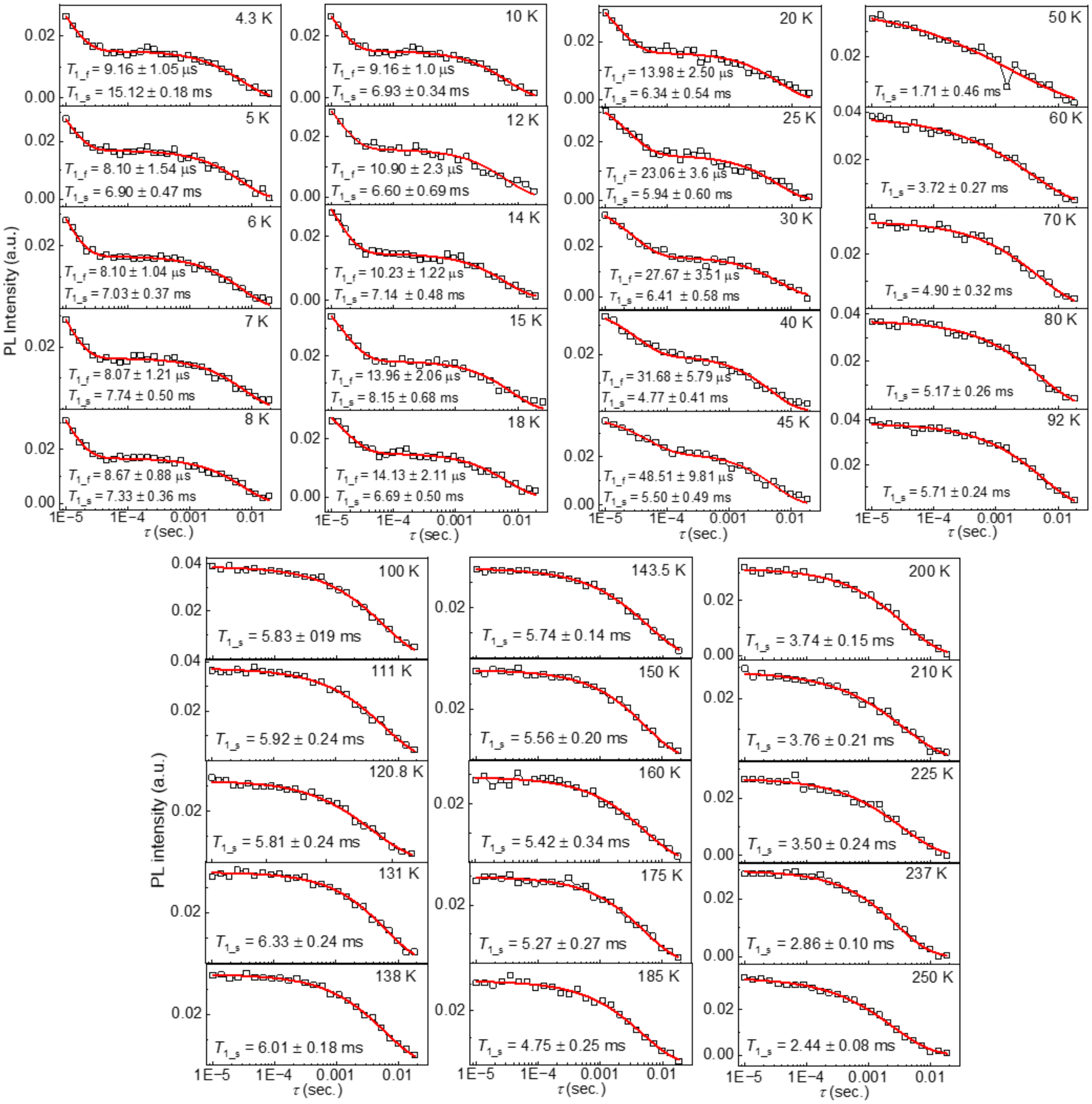

**Figure S7:** Measured (open squares) $T_1$ acquired beneath flake-2 at different temperatures at 9.59 mT. The Solid red lines represent fits of the measured $T_1$ curves using the stretched-exponential decay function described by equation **(5)** and the double-exponential decay function described by equation **(6)** discussed in the main text.

**Comparison of NV-$T_1$ beneath α−$RuCl_3$ flakes, flake-1 and flake-2:** To investigate the thickness dependence of the NV's $T_1$, we compare the temperature-dependent $T_1$ values measured beneath flake-1 (~8 nm thick) and flake-2 (~95 nm thick), as shown in Figure S8. The $T_1$ times were extracted from the fitting procedures described in the main text and are plotted on a logarithmic scale to facilitate comparison over the full temperature range.

A clear thickness-dependent effect is observed, with shorter $T_1$ times beneath the thicker flake-2 compared with the thinner flake-1. This behaviour suggests an enhanced magnetic noise contribution from the thicker α−$RuCl_3$ layer at the NV transition frequencies[43]. Both flakes exhibit a pronounced decrease in $T_1$ upon cooling below ~100 K, indicating an increase in the magnetic noise experienced by the NV centers over the corresponding temperature range. Furthermore, the relaxation beneath flake-2 exhibits two distinct decay rates (fast and slow) below 50 K, highlighting a more complex relaxation behaviour in this temperature regime discussed in the main text.

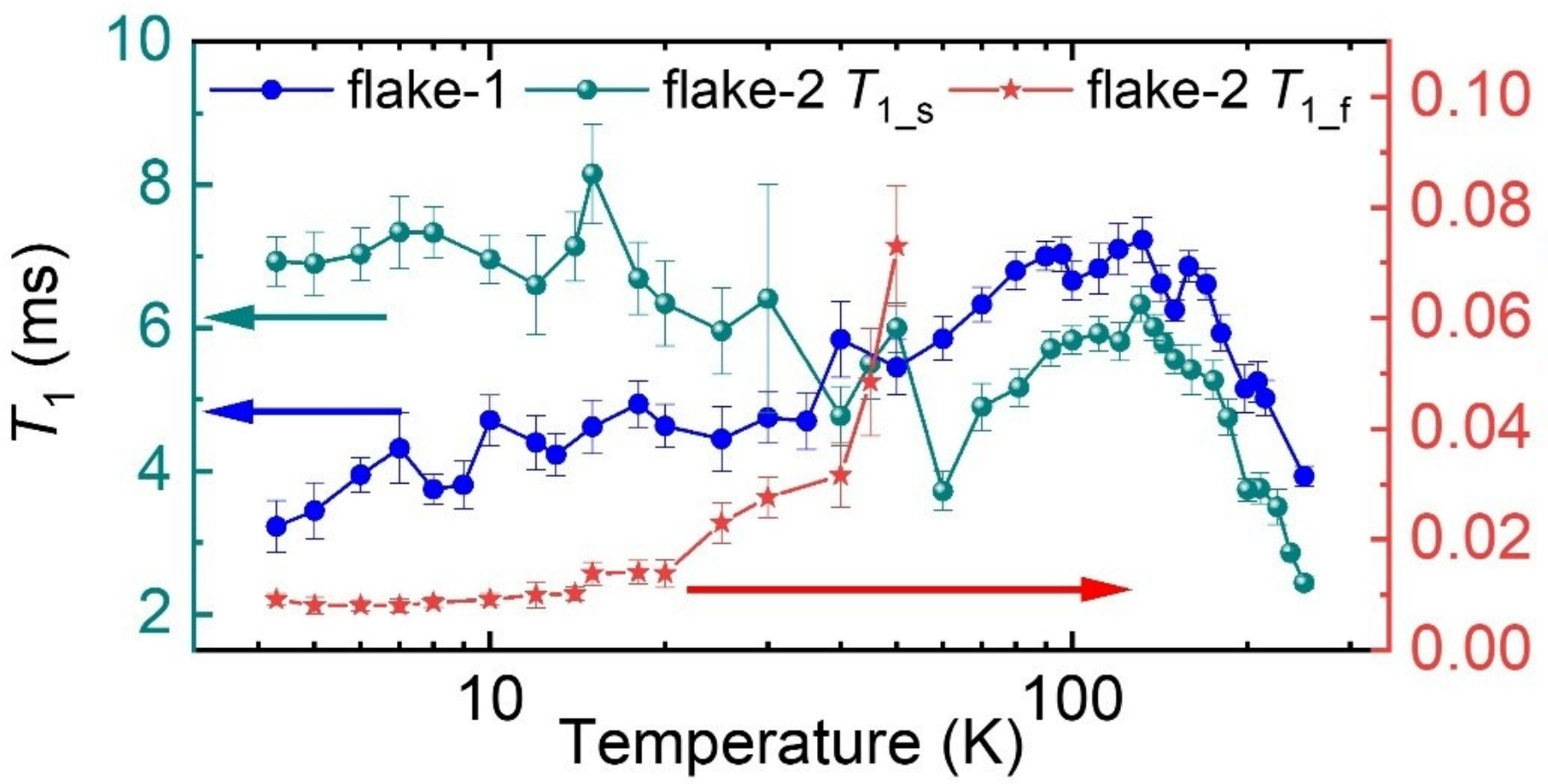

**Figure S8**: Comparison of the NV-$T_1$ relaxation times measured beneath flake-1 (~8 nm) and flake-2 (~95 nm) as a function of temperature at an external magnetic field of 9.59 mT. For flake-2, the fast and slow decay components are plotted on the right and left y-axes, respectively. The arrows in the figure highlight the $T_1$ axis for slow and fast decays.

**$T_1$ data under α−$RuCl_3$ flake-3 at 9.59 mT:** We measured the temperature dependence of the NV $T_1$ of flake-3 over a temperature range of 4.3 – 250 K at 9.59 mT, as shown in Figure S9. The experimental data are well described by a single stretched-exponential decay function over the temperature range of 50 – 250 K. Below 50 K, however, a single-exponential decay model is no longer adequate, and the data are better described by a double-exponential decay function over the temperature range of 4.3 – 45 K. The extracted $T_1$ times are presented in Figure S10a, revealing a clear change in the temperature-dependent relaxation behavior around 100 K, followed by a splitting into fast and slow relaxation components below 50 K. This behavior is similar to that observed beneath flake-2 presented in the main text.

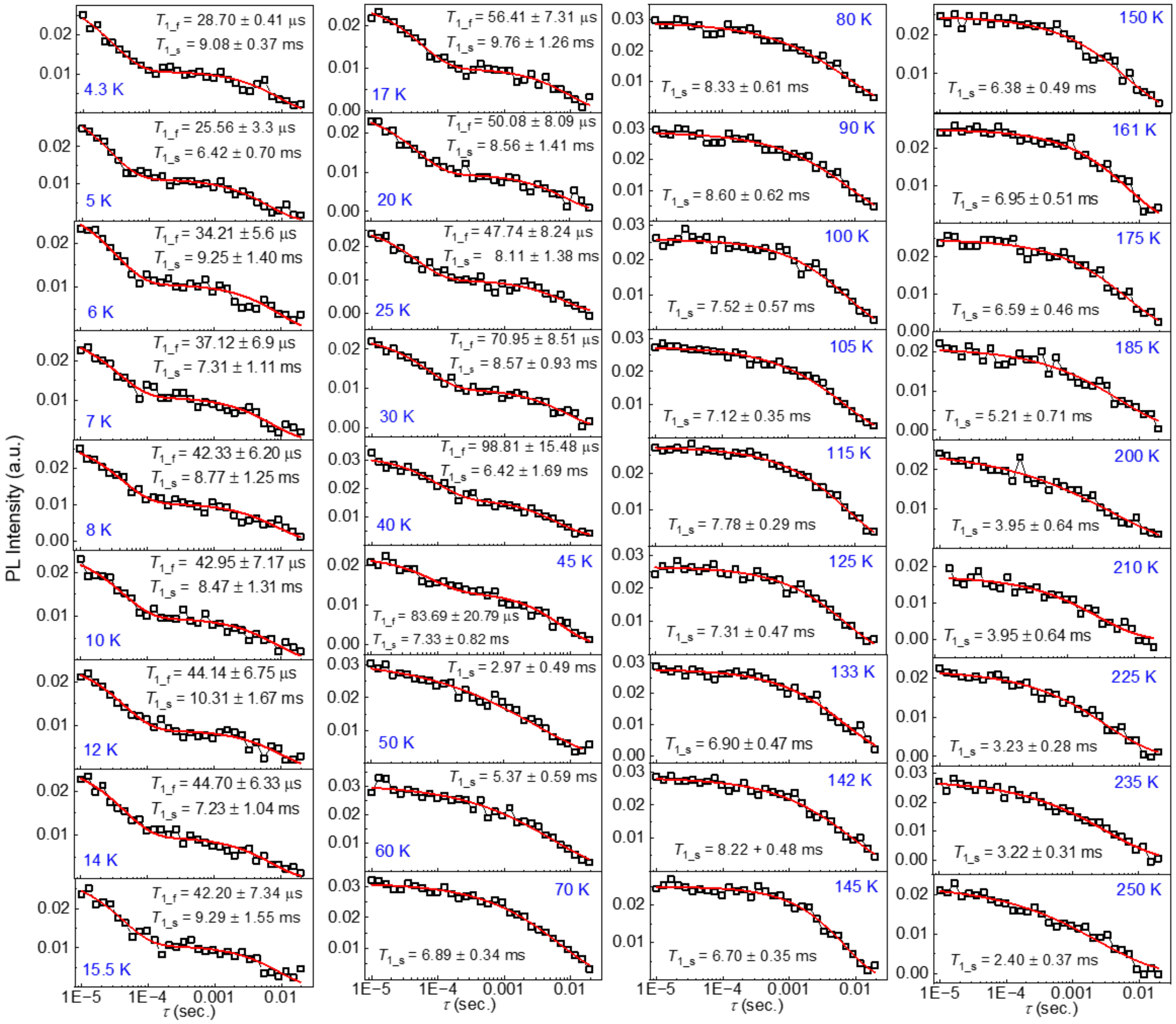

**Figure S9:** Measured (open squares) $T_1$ acquired beneath flake-3 at different temperatures at 9.59 mT. The Solid red lines represent fits of the measured $T_1$ curves using the exponential decay functions described by Equations **(5)** and **(6)** in the main text.

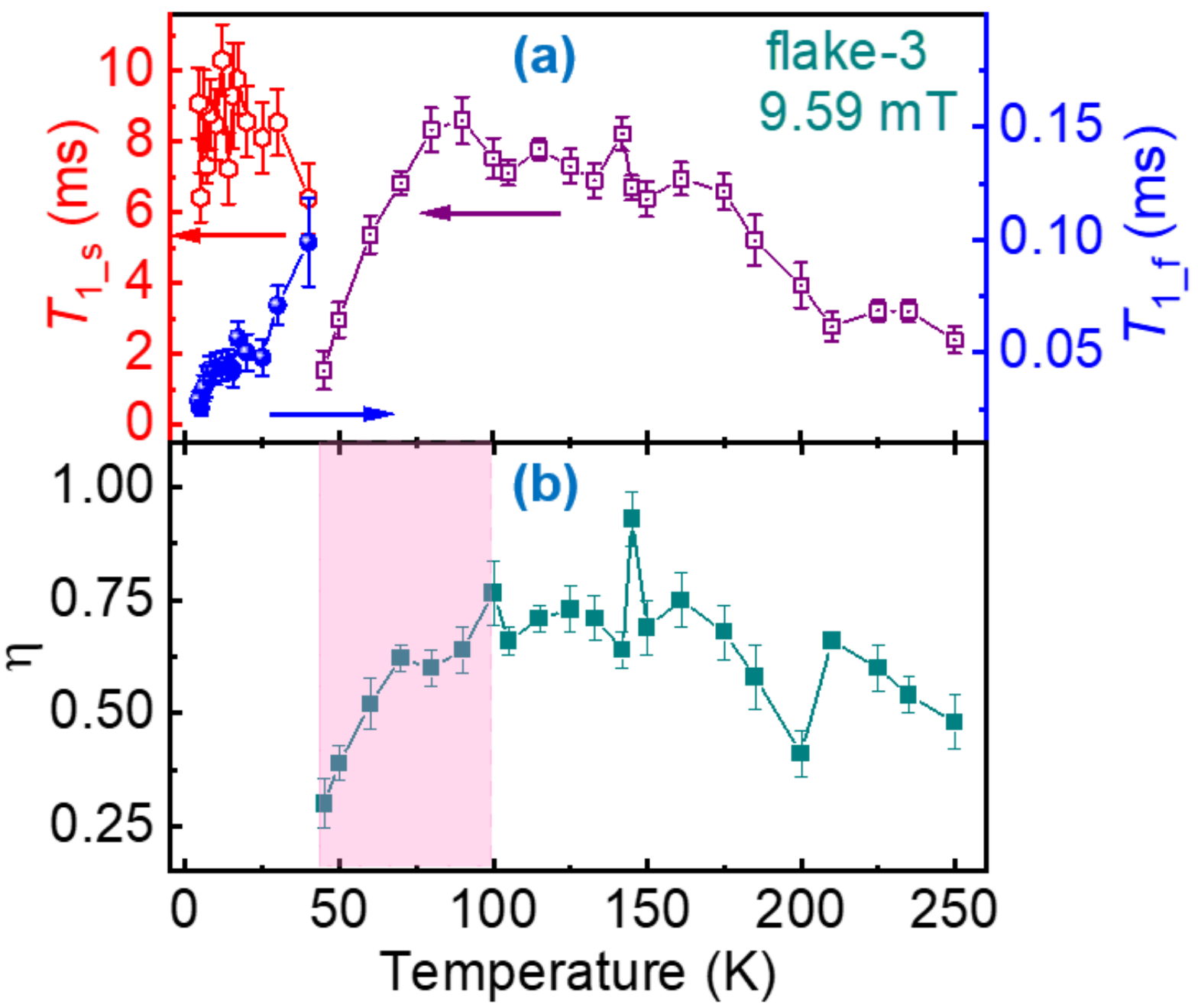

**Figure S10**: (a) Temperature-dependent NV $T_1$ beneath flake-3. Below 45 K, the fast and slow decay components are shown in the right and left panels, respectively. (b) Temperature dependence of the stretched-exponential parameter η. The shaded region shows a sharp change in η below 100 K. The arrows in (a) highlight the $T_1$ axis for slow and fast decays.

**$T_1$ data under α−$RuCl_3$ flake-3 at 22.66 mT:** We measured the temperature dependence of the NV $T_1$ of flake-3 (~150 nm) over a temperature range of 4.3 – 250 K at 22.66 mT, as shown in Figure S11. The experimental data are well fitted using a single stretched-exponential decay function over the temperature range of 50 – 250 K and a double-exponential decay function from 4.3 to 45 K. The extracted $T_1$ and corresponding stretching parameters are presented in Figure 3 of the main text.

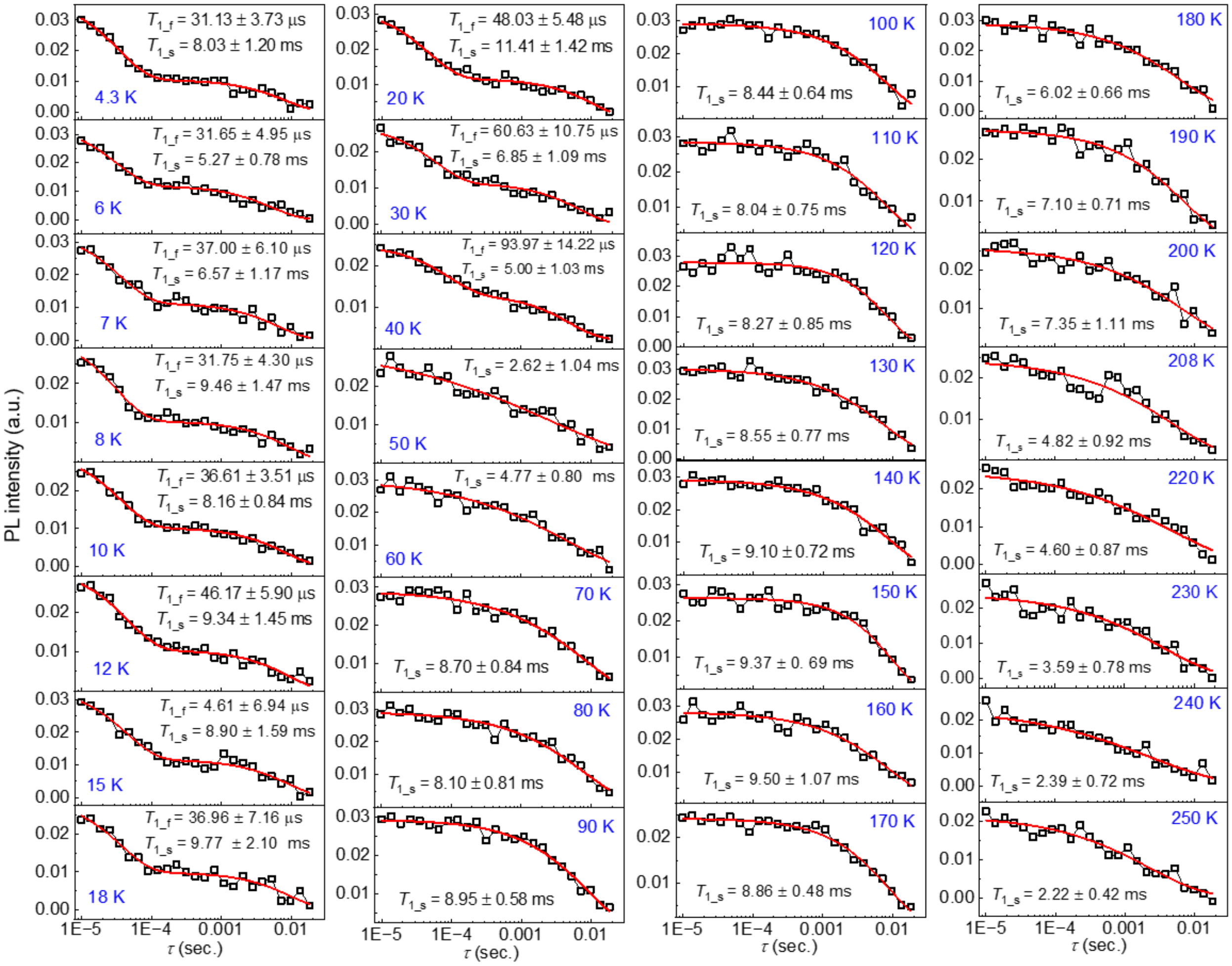

**Figure S11:** NV relaxometry measurements acquired beneath flake-3 at an external magnetic field of 22.66 mT over a temperature range of 4.3–250 K. Solid red lines represent fits to the experimental data using a single stretched-exponential decay function from 50 to 250 K and a double-exponential decay function from 4.3 to 45 K, as described by Equations (5) and (6), respectively, in the main text.